\documentclass[aps,prb,twocolumn,superscriptaddress,floatfix,longbibliography,raggedbottom]{revtex4-2}

\usepackage{amsmath,amssymb} % math symbols
\usepackage{bm} % bold math font
\usepackage{graphicx} % for figures
\usepackage{comment} % allows block comments
\usepackage{textcomp} % This package is just to give the text quote '
\usepackage[nolist,nohyperlinks]{acronym}
\usepackage{float}
\usepackage{upgreek}
\usepackage{siunitx}
\usepackage{enumitem}
\setlist{noitemsep,leftmargin=*,topsep=0pt,parsep=0pt}

\usepackage{xcolor} % \textcolor{red}{text} will be red for notes
\definecolor{lightgray}{gray}{0.6}
\definecolor{medgray}{gray}{0.4}

\usepackage{hyperref}
\hypersetup{
colorlinks=true,
urlcolor= blue,
citecolor=blue,
linkcolor= blue,
}

\newif\ifptitle
\newif\ifpnumber
\newcounter{para}

\ptitletrue  % comment this line to hide paragraph titles
\pnumbertrue  % comment this line to hide paragraph numbers

\newcommand{\mytitle}{Impact of higher-order dispersion on frequency-modulated combs}

\begin{document}

\title{\mytitle}

\author{Nikola Opa$\mathrm{\check{c}}$ak}
\email[]{nikola.opacak@tuwien.ac.at}
\affiliation{Institute of Solid State Electronics, TU Wien, Vienna, Austria}
%\affiliation{Harvard John A. Paulson School of Engineering and Applied Sciences, Harvard University, Cambridge, MA 02138, USA}

\author{Barbara Schneider}
\affiliation{Institute for Quantum Electronics, ETH Zurich, Zurich, Switzerland}

\author{J\'{e}r\^{o}me Faist}
\affiliation{Institute for Quantum Electronics, ETH Zurich, Zurich, Switzerland}

\author{Benedikt Schwarz}
 \email[]{benedikt.schwarz@tuwien.ac.at}
\affiliation{Institute of Solid State Electronics, TU Wien, Vienna, Austria}
%\affiliation{Harvard John A. Paulson School of Engineering and Applied Sciences, Harvard University, Cambridge, MA 02138, USA}

%\date{\today}

\begin{abstract}

Frequency-modulated (FM) combs form spontaneously in free-running semiconductor lasers and possess a vast potential for spectroscopic applications.
Despite recent progress in obtaining a conclusive theoretical description, experimental FM combs often exhibit non-ideal traits, which prevents their widespread use.
Here we explain this by providing a clear theoretical and experimental study of the impact of the higher-order dispersion on FM combs.
We reveal that spectrally-dependent dispersion is detrimental for comb performance and leads to a decreased comb bandwidth and the appearance of spectral holes. These undesirable traits can be mended by applying a radio-frequency modulation of the laser bias. We show that electrical injection-locking of the laser leads to a significant increase of the comb bandwidth, a uniform-like spectral amplitudes, and the rectification of the instantaneous frequency to recover a nearly linear frequency chirp of FM combs.

\end{abstract}

\maketitle

Perfectly periodic waveforms of light, known as optical frequency combs, stand as one of the pillars of modern optics, with applications ranging from fundamental science to precise frequency metrology~\cite{diddams2020optical}. 
Recent years have witnessed to the enormous efforts behind decreasing the footprint of comb solutions -- inciting rapid progress of chip-scale integrated comb generators. Among these, semiconductor \ac{fp} lasers are of large importance as they are compact and possess substantial broadband gain provided by electrical pumping.
The research interest in semiconductor lasers has recently particularly peaked, largely owing to their capacity to emit the so-called \ac{fm} combs.
%and have recently captured wide attention of researchers due to their ability to emit \ac{fm} combs. 
These combs constitute an exciting novel alternative to generate equidistant comb spectra, reported to date in various semiconductor laser types such as the \ac{qcl}~\cite{hugi2012mid,burghoff2014terahertz,singleton2018evidence}, interband cascade laser~\cite{schwarz2019monolithic}, quantum dot laser~\cite{hillbrand2020inphase}, VCSEL~\cite{kriso2021signatures}, and quantum well laser diode~\cite{day2020simple,sterczewski2020frequency,sterczewski2022battery}.
The traditional and well-established \ac{am} combs, comprising a train of short periodic light pulses emitted from mode-locked lasers~\cite{haus2000mode}, often rely on a series of external optical elements to form, potentially requiring tabletop-sized setups. In sharp contrast, \ac{fm} combs form spontaneously in semiconductor \ac{fp} lasers without the need of any additional optical elements, which makes them especially appealing for integrated applications.
\ac{fm} combs stand out all the more as they are not characterized by pulses, but rather a constant intensity where the instantaneous frequency is modulated instead with a periodic linear chirp.

\begin{figure}[h!]
	\centering
	\includegraphics[width = 1\columnwidth]{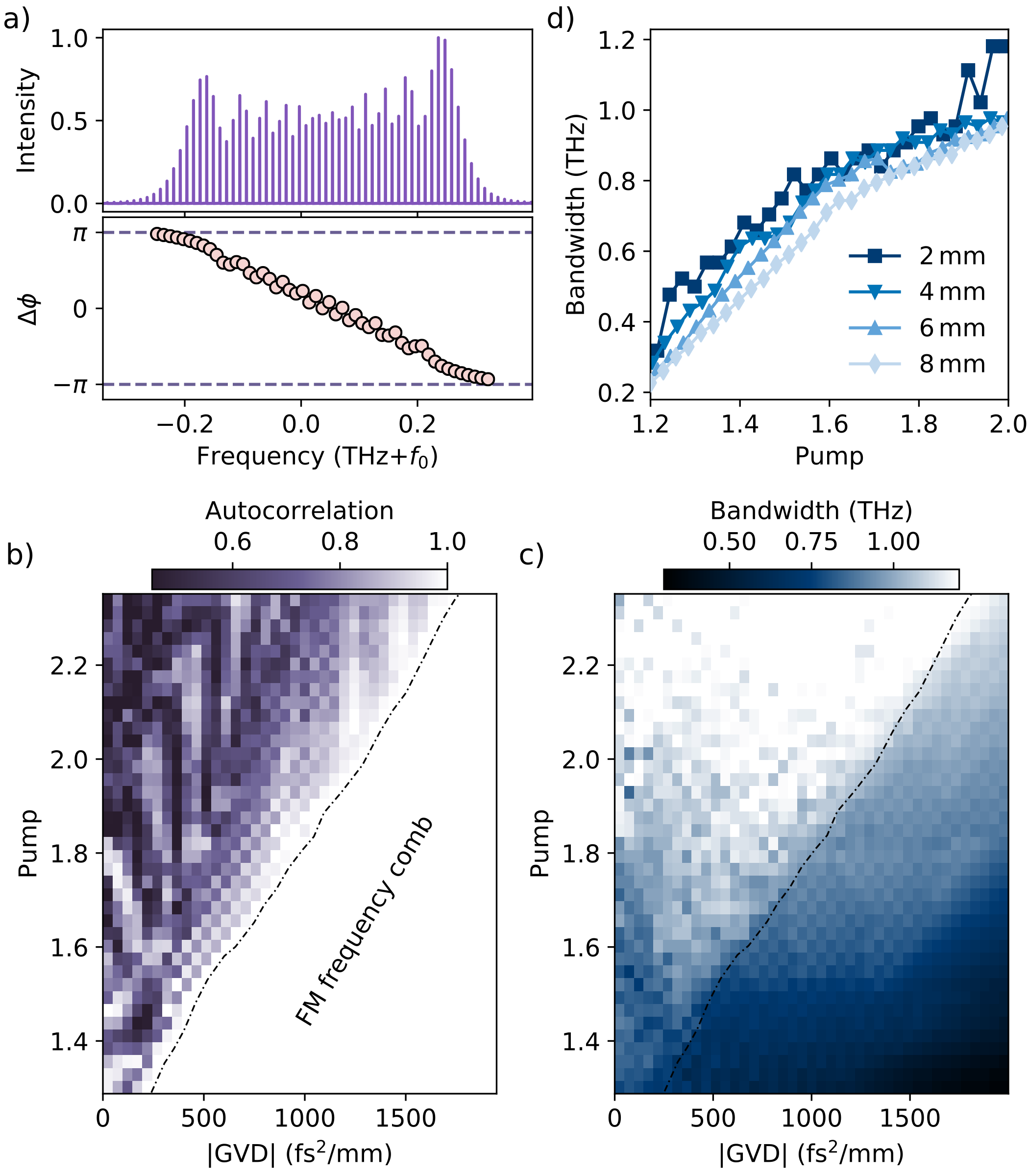}
	\caption{ \textbf{Simulated FM frequency combs induced by constant GVD.} \textbf{(a)} The intensity spectrum (top) and the intermodal phases (bottom) of a comb obtained for GVD=1500$\,\mathrm{fs}^2/\mathrm{mm}$. \textbf{(b)} Calculated autocorrelation as both the GVD and the laser pump are swept. The sign of the GVD only affects the direction of the frequency chirp. The pump is defined as $(J-J_{tr})/(J_{th}-J_{tr})$, where $J$ is the laser current density, $J_{th}$ is the lasing threshold, and $J_{tr}$ is the transparency current. Lasing occurs for pump values larger than 1. FM combs are obtained in the indicated parameter space, separated with a dashed-dotted line from the unlocked states. Each pixel represents a state obtained after simulated 20 000 cavity roundtrips. \textbf{(c)} Obtained spectral width for the same parameter sweep. Optimal GVD which yields the largest comb bandwidth coincides with the minimum dispersion necessary for stable comb formation given with the dashed-dotted black line. \textbf{(d)} Comb bandwidth at a fixed GVD value as the pump is swept for different FP cavity lengths.   }
	\label{fig1}
\end{figure}

The physical origin of \ac{fm} combs was the topic of exhaustive theoretical studies in recent years. Initial studies revealed the crucial roles of \ac{shb} and \ac{fwm} in the formation of equidistant multimode spectra~\cite{khurgin2014coherent,mansuripur2016single,dong2018physics}. However, the particular modal phase arrangement of \ac{fm} combs which leads to their strikingly conspicuous linear frequency chirp originates purely from a Kerr third-order optical nonlinearity or a \ac{gvd} present in the laser system with finite facet reflectivities~\cite{opacak2019theory,opacak2021frequency,silvestri2020coherent,burghoff2020unraveling,humbard2022analytical}. 
The latter was particularly shocking, as even a small \ac{gvd} was believed to be detrimental for coherent laser operation, as in the case of \ac{am} combs. Despite this notion, a finite \ac{gvd} was shown to not only be potentially necessary for coherent comb emission, but also to increase the optical bandwidth of the comb~\cite{beiser2021engineering}.
Regardless of the evident breakthroughs in theoretical understanding, experimental \ac{fm} combs often continue to be plagued by poor performance exhibiting nonuniform spectra and low optical bandwidth, thus limiting their use.

In this work, we present a combined experimental and theoretical study of the influence of higher-order dispersion on \ac{fm} combs.
Although the role of higher-order dispersion is well understood in the formation of other comb types e.g. Kerr combs~\cite{bao2017high}, where it is even utilized to create octave-spanning spectra~\cite{pfeiffer2017octave}, the impact it has on \ac{fm} combs remained unclear.
We demonstrate that a spectrally-dependent dispersion has a severe impact on \ac{fm} comb performance. Increasing the higher-order (third) contribution leads to poor experimentally-observed characteristics such as nonuniform comb spectra, nonlinear frequency chirp, and spectral narrowing -- which are undesirable for applications.
Lastly, we show that even in the case of high third-order dispersion, the negative impact on \ac{fm} combs can be diminished and even reversed by applying a \ac{rf} modulation of the laser bias. The resulting injection-locking of the laser yields the highly-desired larger comb bandwidth, a uniform optical spectrum, and recovers the linear-like frequency chirp.

\begin{figure}[t]
	\centering
	\includegraphics[width = 1\columnwidth]{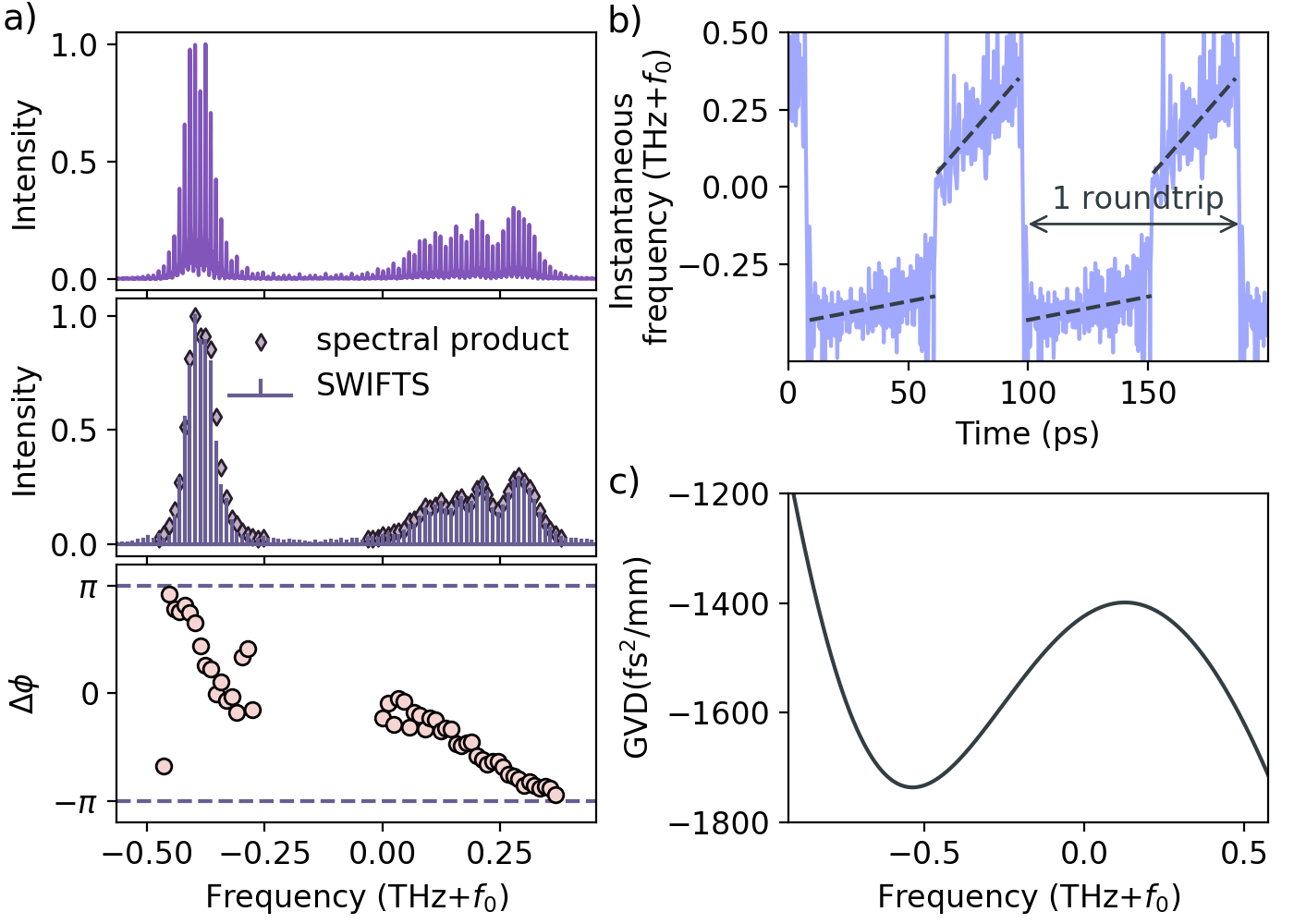}
	\caption{ \textbf{Experimentally-measured FM comb with a large higher-order dispersion.} \textbf{(a)} Intensity spectrum exhibiting nonuniform amplitude distribution and a spectral hole in the middle (top);  SWIFTS spectrum (middle) matches the geometric average of neighboring intensity spectrum amplitudes, indicating comb operation; intermodal phases (bottom). \textbf{(b)} Piecewise-linear instantaneous frequency, indicated with the dashed lines. The discontinuity occurs due to the hole in the middle of the comb spectrum. \textbf{(c)} Measured subthreshold GVD contains linear and higher-order contributions. The measurement was taken at the bias of 0.59$\,A$, just below the lasing threshold.}
	\label{fig2}
\end{figure}

To theoretically study the behavior of \ac{fm} combs, we employ spatio-temporal numerical simulations of the master equation, derived from the Maxwell-Bloch system~\cite{opacak2019theory,opacak2021frequency}. A typical theoretical \ac{fm} comb, obtained with a second-order (constant) dispersion of 1500$\,\mathrm{fs}^2/\mathrm{mm}$ in a 4$\,\mathrm{mm}$ long \ac{fp} cavity is displayed in Fig.~\ref{fig1}a). It exhibits an ideal uniform-like spectrum with the distinct linear intermodal phases that cover the full spectral range of $2\pi$, corresponding to a linear frequency chirp and a quasi-constant intensity output. In the absence of Kerr nonlinearity, \ac{gvd} provides the only mechanism that shapes the linear chirp of \ac{fm} combs in \ac{fp} lasers with partially reflective facets. Fig.~\ref{fig1}b) displays the autocorrelation value of simulated laser states, as the second-order dispersion and the laser pump are swept, where lasing occurs for pump values larger than 1. The autocorrelation is calculated between the emitted field over one roundtrip and the emitted field delayed by 500 roundtrips. It is evident that an \ac{fm} comb, characterized by the autocorrelation of 1 (smaller values indicate an unlocked state), is obtained only for a nonzero value of the \ac{gvd}. Furthermore, the minimum value of \ac{gvd} required for stable comb formation increases with the laser pump (dashed-dotted line). Potential presence of a nonzero Kerr nonlinearity would shift the FM comb parameter space horizontally along the  GVD-axis~\cite{opacak2019theory,beiser2021engineering,opacak2021frequency,burghoff2020unraveling}.
Not all obtained comb states possess equally appealing characteristics, which is apparent from Fig.~\ref{fig1}c), where we display the comb bandwidth.
Aiming for the largest spectral width, the laser should  be operated at a high bias point, and have a dispersion that is just large enough to obtain stable comb operation, but not larger. The influence of the \ac{fp} cavity length is visible in Fig.~\ref{fig1}d), where we have swept the laser pump for 4 different lengths while keeping the \ac{gvd} constant. While strongly dependent on the laser bias, the comb bandwidth increases for shorter cavities, at a fixed pump. Identical qualitative behavior was obtained
by following analytic theory~\cite{burghoff2020unraveling,humbard2022analytical}, although the predicted dependence on the cavity length was stronger, probably due to neglecting of the finite laser gainwidth in the analytic approach. 

 \begin{figure*}[t]
	\centering
	\includegraphics[width = 1\textwidth]{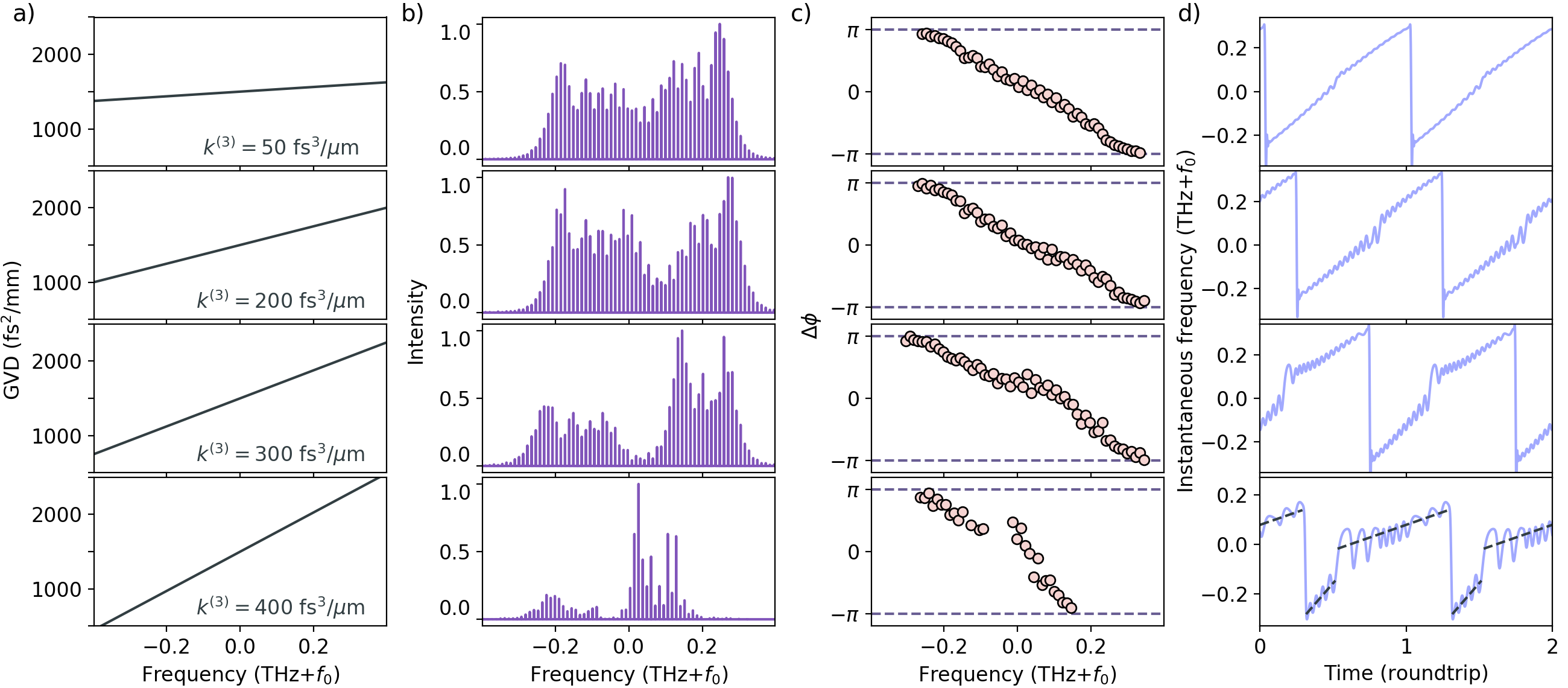}
	\caption{ \textbf{Influence of the third-order dispersion on FM combs in numerical simulations.}  \textbf{(a)} Total spectrally-dependent GVD$(\omega)$ with an increasing third-order (linear) dispersion $k^{(3)}$. Fourth- and higher-order contributions are omitted for the sake of numerical stability. The evolution of the 
 \textbf{(b)} intensity spectra, \textbf{(c)} intermodal phases, and the \textbf{(d)} instantaneous frequency as the third-order dispersion is increased. }
	\label{fig3}
\end{figure*}

Unlike the ideal theoretical \ac{fm} combs, experimentally-obtained results are often riddled with undesired traits such as nonuniform spectra with holes, as shown in the top of Fig.~\ref{fig2}a), measured for a 4$\,\mathrm{mm}$ long \ac{fp} \ac{qcl}. More details on the laser active region and cavity design can be found in~\cite{schneider2021controlling}. Employing \ac{swifts}~\cite{burghoff2015evaluating} we can prove the coherence of the comb state and extract the intermodal phases by recording interferograms at the comb roundtrip frequency. Due to the hole in the middle of the intensity spectrum, the intermodal phases arrange themselves in two distinct linear patterns, together covering the whole range of $2\pi$. As a consequence, the instantaneous frequency is a piecewise linear function during one cavity roundtrip (Fig.~\ref{fig2}b)). The experimental FM comb state does not resemble much the simulated ideal state in Fig.~\ref{fig1}, urging to discover the cause behind the observed degradation. 
Fig.~\ref{fig2}c) displays the measured subthreshold dispersion~\cite{hofstetter1999measurement}, showing non-constant values with a significant linear contribution around the lasing range,
corresponding to a large third-order dispersion. This is in sharp contrast with the simulations in Fig.~\ref{fig1} where constant second-order dispersion values were used, already pointing that the probable culprit behind nonuniform experimental spectra is the higher-order dispersion.

To corroborate this hypothesis, we performed numerical simulations of the master equation incorporating the third-order dispersion $k^{(3)}$, to account just for the linear dependence of the dispersion observed in Fig.~\ref{fig2}d). This contribution is sufficient to explain the experimental behavior. The total dispersion is then GVD$(\omega)=k^{(2)}+k^{(3)}(\omega-\omega_0)$, where $k^{(2)}$ is the second-order dispersion, and $\omega$ is the optical angular frequency. Mathematically, the additional term in the master equation is introduced as a third-order temporal derivative of the light field~\cite{opacak2022thesis}. Terms $\mathcal{O}(k^{(4)})$ describe nonlinear dispersion dependence $\mathcal{O}(\omega^{(2)})$ and are omitted as they would introduce higher-order derivates and issues with numerical stability. The origin of $k^{(3)}$ is purely considered to be from the cavity and material dispersion. An additional contribution is due to the gain spectral profile itself and is expected to increase in importance for narrower gainwidths or inhomogenous gain broadening e.g. in heterogenous \acp{qcl}~\cite{bachmann2016dispersion}.
Column a) in Fig.~\ref{fig3} shows the gradual increase of the the third-order dispersion which was inserted in numerical simulations to study its impact on the comb state shown in Fig.~\ref{fig1}a). Columns b), c), and d) in Fig.~\ref{fig3} display the resulting comb spectra, intermodal phases, and the instantaneous frequency, respectively. Although maintaining comb operation, the intensity spectra progressively develop an irregular amplitude distribution, resembling experimental comb spectra that are often found in literature~\cite{hugi2012mid,burghoff2014terahertz,singleton2018evidence,hillbrand2020inphase,schwarz2019monolithic,day2020simple,sterczewski2022battery,kriso2021signatures,sterczewski2020frequency}. At the same time, the intermodal phases and the instantaneous frequency increasingly deviate from a linear pattern. For the highest value of third-order dispersion ($k^{(3)}=400\,\mathrm{fs}^2/\upmu m$), the spectrum splits in two separated lobes, alike in Fig.~\ref{fig2}a). Further resemblance to the experimental state is observed from the intermodal phases and the instantaneous frequency, which in approximation becomes a piecewise linear function during one roundtrip.

\begin{figure}[t]
	\centering
	\includegraphics[width = 1\columnwidth]{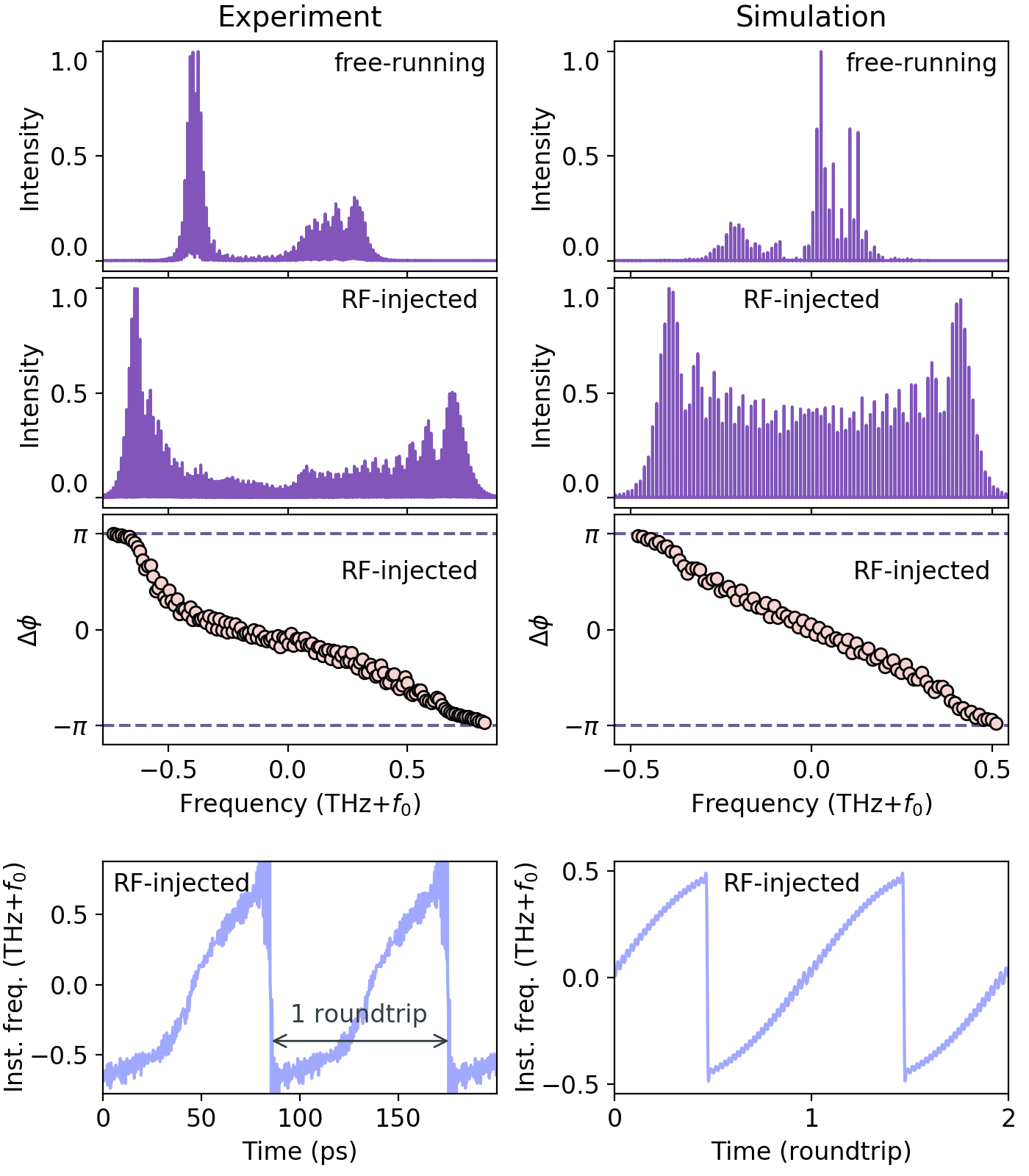}
	\caption{ \textbf{The effect of RF modulation of the laser bias.} Experimental data (left) and numerical simulations (right) indicate that modulating the laser bias leads to the broadening of the spectrum and flattening of the spectral amplitudes. The intermodal phases and the instantaneous frequency restore nearly linear behavior characteristic to ideal FM combs. }
	\label{fig4}
\end{figure}

In order to reach the full potential of \ac{fm} combs for widespread applications, it is of crucial importance to achieve reproducible high-quality comb behavior, as in the numerical simulations. First of all, having a broadband uniform comb spectra is of high interest for spectroscopy. On the other hand, obtaining a linear frequency chirp allows the use of a pulse compressor via group delay compensation~\cite{singleton2019pulses} to achieve femtosecond pulse emission~\cite{taeschler2021femtosecond}.  
In Fig.~\ref{fig4} we demonstrate, both experimentally and theoretically, how to recover these highly desired traits even in \ac{fm} combs that possess a large higher-order dispersion by employing an \ac{rf} modulation of the laser bias. Modulating the bias of a semiconductor laser at the roundtrip frequency enables coherent control of many of its characteristics -- which was so far successfully used to injection-lock the comb beatnote~\cite{hillbrand2018coherent,schwarz2019monolithic,hillbrand2020inphase}, eliminate higher-order transverse cavity modes~\cite{dalcin2022coherent}, emit actively mode-locked pulses~\cite{hillbrand2020mode}, and increase the spectral bandwidth of \ac{fm} combs~\cite{beiser2021engineering,schneider2021controlling}. The modulation should be done on a short end-section of the \ac{fp} cavity to maximize the effect~\cite{piccardo2018time}.
In the top of Fig.~\ref{fig4} we show free-running experimental and simulated \ac{fm} comb states with large higher-order dispersion, taken from Fig.\ref{fig2}a) and bottom of Fig.~\ref{fig3}b), respectively. The impact of the \ac{rf} injection is visible from the panels below, showing agreement between the experimental and simulation results. 
The comb spectra become more uniform and the spectral hole disappears. Apart from this, another striking effect is reflected in a significant broadening of the comb bandwidth of around 100$\,$\%~\cite{beiser2021engineering,schneider2021controlling}. In the theoretical model, the modulation is directly implemented to the laser bias of a short section covering 10$\,$\% of the cavity as $J(t)=J_{\mathrm{DC}}+J_{\mathrm{m}}\cos(\omega_{\mathrm{m}}t)$, where $\omega_{\mathrm{m}}$ is the modulation frequency and we set $J_{\mathrm{m}}=0.05J_{\mathrm{DC}}$~\cite{hillbrand2020mode}.
In the experiment, a 4 mm Fabry-P\'erot QCL with a microstrip-like line geometry~\cite{schneider2021controlling} was used. Both, a DC-bias of 950 mA and a 25 dBm AC bias modulation of 11.090 GHz, which matches the free-running intermode beating, were supplied using a high-power coplanar RF-probe. Whereas the device has no dedicated modulation section, the point of contact with the probe is very close to the front-facet, which has a similar effect as modulating a short end-section.
The impact of bias modulation is visible from the intermodal phases as well, which splay over $2\pi$ in a single nearly linear pattern. Slightly worse behavior in experiments could be explained by the existence of finite $\mathcal{O}(k^{(4)})$ terms. The piecewise linear instantaneous frequency of free-running combs becomes significantly rectified, thus enabling the use of group delay compensation schemes to achieve pulse compression.

In this work, we provided a study of the impact of higher-order dispersion on \ac{fm} combs by utilizing both numerical simulations of a spatio-temporal theoretical model and experimental measurements.
We reveal that the presence of finite higher-order dispersion $\mathcal{O}(k^{(3)})$ negatively affects \ac{fm} comb characteristics by lowering the spectral bandwidth, producing spectral holes in the spectrum, and resulting in a nonlinear frequency chirp.
In order to boost \ac{fm} comb uses for broadband spectroscopic applications, it is therefore crucial to mitigate higher-order dispersion if possible. One way of achieving this would be to tailor the spectral gain profile or change the laser cavity in an effort to make the \ac{gvd} constant within the lasing range. 
Here we demonstrated a simple technique which achieves this and does not rely on any modifications of the existing laser system.
By modulating the laser current around the roundtrip frequency we overcome the severe limitations imposed by the higher-order dispersion by doubling the comb bandwidth, flattening the intensity spectrum and recovering a nearly linear frequency chirp. The further use of \ac{rf} modulation could enhance the use of \ac{fm} combs for dual-comb spectroscopy.

\section*{Acknowledgements}
N.~O. and B.~Schwarz have received funding from the European Research Council (ERC) under the European Union’s Horizon 2020 research and innovation programme (Grant agreement No. 853014). B. Schneider and J. F. have received funding from the European Research Council (ERC) under the European Union’s Horizon 2020 research and innovation programme (Grant agreement No. 820419) and from the Swiss Innovation Agency Innosuisse  (Grant agreement No. 2155008433).
% \todo{ETH: Please list any funding agencies if needed.}

\begin{acronym}

\acro{fm}[FM]{frequency-modulated}
\acro{am}[AM]{amplitude-modulated}
\acro{qcl}[QCL]{quantum cascade laser}
\acro{fp}[FP]{Fabry-P\'{e}rot}
\acro{shb}[SHB]{spatial hole burning}
\acro{gvd}[GVD]{group velocity dispersion}
\acro{lef}[LEF]{linewidth enhancement factor}
\acro{fwm}[FWM]{four-wave mixing}
\acro{rf}[RF]{radio-frequency}
\acro{swifts}[SWIFTS]{Shifted-Wave Interference Fourier Transform
Spectroscopy}

\end{acronym}

\end{document}